\documentclass[aps, prl, twocolumn]{revtex4-1}
\usepackage{graphicx}% Include figure files
\usepackage{float}
\usepackage{amsmath}
\usepackage{amsfonts}
\usepackage{color}

\begin{document}

\title{Emergence of 6-particle ``hexciton'' states in WS$_2$ and MoSe$_2$ monolayers}

\author{J. Choi$^{1,2}$, J. Li$^{1,3}$, D. Van Tuan$^{4}$, H. Dery$^{4,5}$, S. A. Crooker$^1$}
\affiliation{$^1$National High Magnetic Field Laboratory, Los Alamos, NM 87545, USA}
\affiliation{$^2$Advanced Instrumentation Institute, Korea Research Institute of Standards and Science, Daejeon 34113, Korea}
\affiliation{$^3$Wuhan National High Magnetic Field Center and School of Physics, Huazhong University of Science and Technology, Hubei 430074, China}
\affiliation{$^4$Department of Electrical and Computer Engineering, University of Rochester, Rochester, New York 14627, USA}
\affiliation{$^5$Department of Physics and Astronomy, University of Rochester, Rochester, New York 14627, USA}

\begin{abstract}
When doped with a high density of mobile charge carriers, monolayer transition-metal dichalcogenide (TMD) semiconductors can host new types of composite many-particle exciton states that do not exist in conventional semiconductors.  Such multi-particle bound states arise when a photoexcited electron-hole pair couples to not just a single Fermi sea that is quantum-mechanically distinguishable (as for the case of conventional charged excitons or trions), but rather couples simultaneously to \textit{multiple} Fermi seas, each having distinct spin and valley quantum numbers. Composite six-particle ``hexciton'' states were recently identified in electron-doped WSe$_2$ monolayers, but under suitable conditions they should also form in \textit{all} other members of the monolayer TMD family. Here we present spectroscopic evidence demonstrating the emergence of many-body hexcitons in charge-tunable WS$_2$ monolayers (at the A-exciton) and MoSe$_2$ monolayers (at the B-exciton). The roles of distinguishability and carrier screening on the stability of hexcitons are discussed.
\end{abstract}

\maketitle

The monolayer transition-metal dichalcogenide (TMD) semiconductors such as WSe$_2$, MoSe$_2$, WS$_2$, and MoS$_2$ host a multitude of excitonic complexes, due in part to the spin-orbit-split nature of their conduction and valence bands at the $K$ and $K'$ valleys of the Brillouin zone \cite{Urbaszek:2018, Muller:2018}. When they are optically allowed and possess a non-zero oscillator strength, these bound complexes manifest as discrete resonances in optical absorption spectra. Early studies of nominally undoped TMD monolayers revealed pronounced absorption peaks from the fundamental electron-hole optical transition (i.e., the $X^0$ neutral exciton) \cite{Splendiani:2010, Mak:2010, Li:2014}. Subsequent studies of charge-tunable TMD monolayers demonstrated the emergence, at lower energy, of additional strong absorption lines when the monolayer was populated with a background Fermi sea of holes or electrons (i.e., the  $X^\pm$ charged excitons) \cite{Ross:2013, Mak:2013, Wang:2017NL}. Whether, and under what conditions, a $X^\pm$ complex is most accurately described as a simple three-particle `trion' (wherein the photoexcited exciton binds a carrier from the Fermi sea \cite{Ross:2013, Mak:2013, Wang:2017NL, Kheng:1993, Astakhov:2002, BarJoseph:2005}), or a four-particle `tetron' (a trion additionally correlated with the resulting hole that is left behind in the Fermi sea \cite{Bronold:2000, Suris:2001, Suris:2003, Combescot:2018, Rana:2020}), or an `exciton-polaron' (an exciton dressed by collective excitations of the Fermi sea \cite{Efimkin:2017, Sidler:2016}), remains an active area of study and discussion \cite{Reichman:2019, Glazov:2020, Rana:2021, Efimkin:2021, Fey:2020, Liu:2021}.  

Regardless of interpretation, all of these descriptions share a common understanding: $X^\pm$ are bound states  arising from the interaction of a photoexcited electron-hole (\textit{e-h}) pair with the \textit{subset} of carriers in the Fermi sea that have \textit{distinguishable} quantum numbers.  In most conventional III-V and II-IV semiconductors such as GaAs or ZnSe, where band extrema occur at the single central $\Gamma$-point valley of the Brillouin zone \cite{Kheng:1993, Astakhov:2002, BarJoseph:2005}, this means that $X^\pm$ forms with mobile carriers having opposite \textit{spin} to that of the photoexcited electron or hole because Pauli exclusion prevents strong short-range interactions with same-spin carriers.  However, the multi-valley nature of monolayer TMDs expands the basis set of available quantum numbers, and band-edge electrons and holes can be distinguished not only by their spin (up or down) but also by their valley degree of freedom ($K$ or $K'$). TMD monolayers therefore permit, under suitable conditions, photoexcited \textit{e-h} pairs to interact with Fermi seas containing \textit{more than one} type of quantum-mechanically distinguishable carrier. As demonstrated recently \cite{Li:2022, vanTuanPRL:2022, vanTuan:2022}, this leads to qualitatively new types of multi-particle composite exciton ground states that can be described as bound six-particle ``hexcitons'' (when photoexcited \textit{e-h} pairs interact with two distinguishable Fermi seas) or even eight-particle ``oxcitons'' (when interacting with three distinguishable Fermi seas). These bound many-body excitonic states have large oscillator strengths and manifest as discrete absorption resonances in linear optical spectroscopy, and emerge at energies even further below $X^0$ and $X^\pm$. We emphasize that these composite hexcitons are optically-allowed ground states of the interacting exciton-Fermi sea system, and are therefore distinct from the many types of optically-forbidden dark excitons and trions that appear only in photoluminescence studies \cite{Urbaszek:2018, Muller:2018, Robert:2017, Malic:2018, He:2020, Yang:2022}, and moreover should not be confused with multi-exciton complexes (such as biexcitons) that appear only at higher photoexcitation intensity \cite{Urbaszek:2018, Muller:2018, Barbone:2018, ZLi:2018, Chen:2018}.

To date, such multi-particle hexcitons have been identified and studied only in electron-doped WSe$_2$ monolayers \cite{Li:2022, vanTuanPRL:2022, vanTuan:2022}. This is due to i) the excellent optical quality of exfoliated WSe$_2$, ii) the ability to electrostatically dope WSe$_2$ to high electron densities, and iii) because composite hexcitons in WSe$_2$ are expected at the low-energy A-exciton optical transitions where spectral linewidths are typically much sharper than at the higher-energy B-exciton.  This latter fact arises from the positive sign of the conduction band (CB) spin-orbit splitting in WSe$_2$ ($\Delta_c >0$) \cite{Song:2013, Kormanyos:2015}, which mandates that A-exciton optical transitions photoexcite electrons to the \textit{upper} CBs in $K$ and $K'$, where Pauli exclusion does not prevent them from interacting with \textit{both} of the distinguishable Fermi seas of electrons that occupy the two lower CBs (as depicted in the band diagram in Fig. 1, one Fermi sea resides in the same valley but has opposite spin, the other has same spin but resides in the opposite valley).   

However, under appropriate conditions, composite hexcitons should also emerge in \textit{all} members of the monolayer TMD family. For example, monolayer WS$_2$ also has a positive $\Delta_c$ and CB structure similar to WSe$_2$, and therefore hexcitons should also appear at its A-exciton under conditions of high electron doping. In contrast, the negative $\Delta_c$ of monolayer MoSe$_2$ \cite{Song:2013, Kormanyos:2015} precludes the existence of hexcitons at its A-exciton (instead, only conventional $X^\pm$ should appear), but \textit{does} allow for the formation of hexcitons at its B-exciton transition.  To date, neither of these predictions have been explicitly tested.  Here, using low-temperature optical absorption measurements of electrostatically-gated WS$_2$ and MoSe$_2$ monolayers, we demonstrate and investigate the emergence of composite hexcitons in both WS$_2$ monolayers (at the A-exciton) and MoSe$_2$ monolayers (at the B-exciton).  Based on the data, the roles of distinguishability and carrier screening on the stability of hexcitons are discussed. 

\begin{figure}[t] 
\center
\includegraphics[width=.45\textwidth]{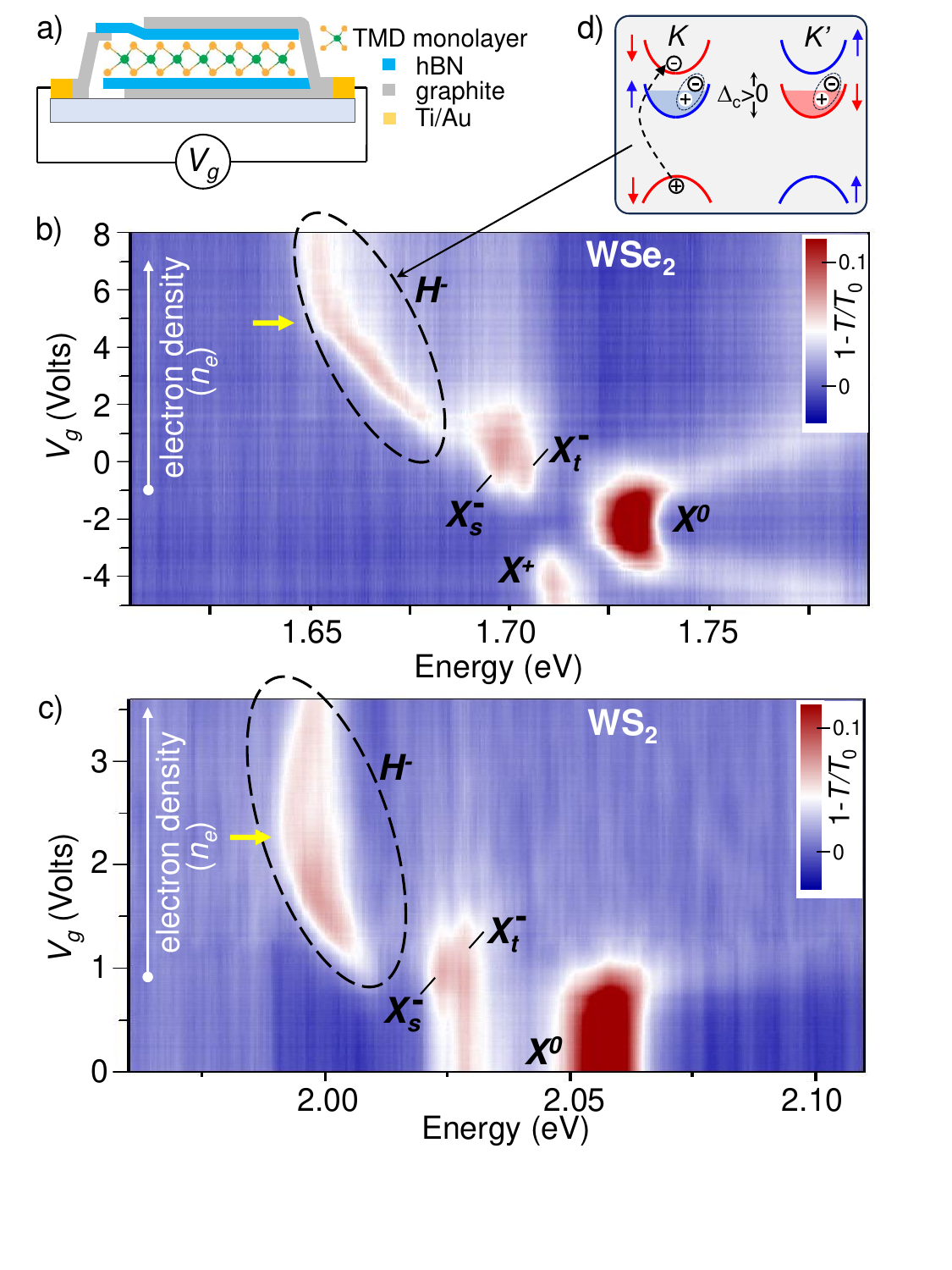}
\caption{a) Schematic of the charge-tunable monolayer TMD structures.  b,c) Gate ($V_g$) dependent optical absorption spectra at 4~K and $B$=0 from monolayer WSe$_2$ and WS$_2$, respectively, at their A-exciton band edges.  Similar to WSe$_2$, WS$_2$ exhibits strong $X^0$ absorption at charge neutrality, conventional $X_{s,t}^-$ absorption at small electron density ($n_e$), and the emergence at even lower energy of a strong ``hexciton'' absorption resonance at larger $n_e$ (dashed oval, labeled as $H^-$).  d) Band diagram of WSe$_2$ and WS$_2$. Extending the picture of $X^\pm$ as 4-particle tetrons, the hexciton corresponds to a 6-particle correlated state that forms when a photoexcited \textit{e-h} pair couples simultaneously to \textit{both} of the quantum-mechanically distinguishable Fermi seas residing in the two lower CBs. This occurs in WSe$_2$ and WS$_2$ at the A-exciton because $\Delta_c >0$, such that optical transitions promote electrons to the upper CBs, where they are distinguishable from electrons in the Fermi sea. Blue/red colors denote spin-up/down bands. For clarity, only optical transitions in the $K$ valley are depicted. The redshift of the hexciton ceases (yellow arrows) when the Fermi sea begins to fill the upper CBs, at $n_e \approx 5 (4) \times 10^{12}$ cm$^{-2}$ for WSe$_2$ (WS$_2$). }
\label{fig1}
\end{figure}

Figure 1a depicts the charge-tunable TMD monolayer samples studied in this work. Single monolayer flakes of WS$_2$, MoSe$_2$, and WSe$_2$ were mechanically exfoliated and sandwiched between thin slabs of hexagonal boron nitride (hBN).  Few-layer graphite flakes were used to electrically contact the monolayer, and to serve as top and bottom gates. In this work, the top and bottom gates were tied together and gate voltage $V_g$ was used to electrostatically dope the monolayers with a background Fermi sea of electrons or holes. Dual gating allows us to attain high carrier densities approaching 10$^{13}$/cm$^2$. Each assembled structure was then positioned and placed directly over the core of a single-mode optical fiber, to ensure a rigid and robust alignment between the optical path and the doped TMD monolayer. This experimental approach \cite{Li:2020, Stier:2018} mitigates the drift and vibration that can otherwise complicate optical studies of TMD monolayers at low temperatures and in the pulsed magnetic fields used in this work. 

The sample-on-fiber assembly was then mounted on a purpose-built probe and loaded into a liquid helium cryostat. Broadband white light from a Xenon lamp was directed down the single mode fiber.  Following transmission through the sample the light passed through a thin-film circular polarizer and was then retro-reflected and directed back into a multimode collection fiber. The transmitted light was dispersed in a 300~mm spectrometer and detected by a charge-coupled device (CCD).  In this way the optical absorption from the doped TMD monolayer was directly measured as $1-T/T_0$, where $T$ is the spectrum of the transmitted light and $T_0$ is a reference spectrum.  We note that absorption spectra typically permit a straightforward evaluation and visualization of exciton oscillator strengths, in comparison to reflectivity studies where lineshapes depend sensitively on interference effects from the surrounding layer structure.

Figure 1b shows a map of the gate-dependent absorption spectra from a WSe$_2$ monolayer at low temperature (4K) and at zero magnetic field, in the spectral range of its A-exciton. When $V_g \approx -2$V, the monolayer is at its charge neutrality point and only the neutral exciton ($X^0$) absorption resonance is observed. At increasingly negative or positive $V_g$, the monolayer becomes lightly doped with mobile holes or electrons, and the well-known $X^\pm$ resonances appear at energies $\approx$20-35 meV below $X^0$.  As extensively described in previous works, $X^\pm$ are optically-active bound states arising from the interaction of the photoexcited \textit{e-h} pair with the carriers in the Fermi sea that possess distinguishable quantum numbers from those of the photoexcited \textit{e-h} pair (i.e., different spin and/or valley). Thus, only a single $X^+$ resonance appears on the hole-doped side, but two conventional $X^-$ resonances appear on the electron-doped side (the so-called singlet $X_s^-$ and triplet $X_t^-$ charged excitons) because there are two distinct Fermi seas with which to interact. The different energies of $X_s^-$ and $X_t^-$ stem from the different amplitude of the short-range electron-hole exchange interaction \cite{Glazov:2020, Hichri:2020, Courtade:2017}.

Most importantly, at higher $n_e$ the $X_s^-$ and $X_t^-$ resonances disappear and a new strong absorption resonance appears at even lower energy ($\approx$15~meV below $X_s^-$), indicating the emergence of a new bound excitonic ground state with large oscillator strength.  First observed in gated WSe$_2$ monolayers in 2013 \cite{Jones:2013} and very clearly resolved in several subsequent studies \cite{Wang:2017NL, Wang:2017, Barbone:2018, SuFeiShi:2020, Liu:2021}, this low-energy absorption resonance -- occasionally called $X^{-\prime}$ in earlier literature -- completely dominates the absorption spectrum of WSe$_2$ at high $n_e$. Moreover, it does not appear in hole-doped WSe$_2$ at the A-exciton, and does not appear in the A-exciton absorption spectrum of gated MoSe$_2$ monolayers \cite{Wang:2017NL, Smolenski:2019}.  Recently, this absorption resonance was identified as a qualitatively new type of many-body composite (six-particle hexciton) state, arising from the simultaneous interaction of the photoexcited \textit{e-h} pair with \textit{both} of the Fermi seas that reside in the lower CBs of monolayer WSe$_2$ \cite{Li:2022, vanTuanPRL:2022, vanTuan:2022}. As depicted in Fig. 1d, each of these two Fermi seas has quantum numbers that are distinguishable from those of the photoexcited electron (one has opposite spin, the other has opposite valley), and -- extending the picture of $X^\pm$ being four-particle tetrons \cite{Suris:2003, Rana:2020} -- the hexciton bound state comprises the photoexcited \textit{e-h} pair, an electron from each of the two distinguishable Fermi seas, and the two Fermi holes that are left behind in the Fermi seas. As described recently, hexcitons are the stable ground state of this interacting exciton-Fermi sea system, and the Fermi holes not only ensure overall charge neutrality but also provide the `glue' that binds the complex \cite{vanTuanPRL:2022, vanTuan:2022}.

\begin{figure}[t]
\center
\includegraphics[width=.45\textwidth]{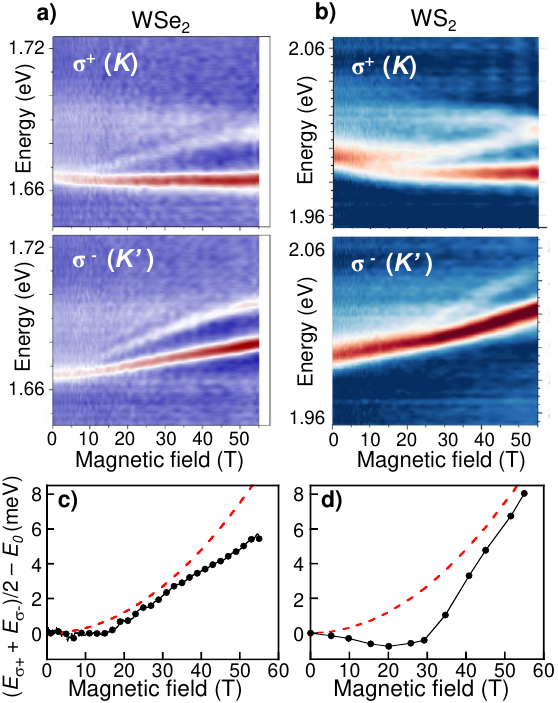}
\caption{a,b) Magnetic field evolution of the hexciton absorption resonance in monolayer WSe$_2$ and WS$_2$, respectively, for both $\sigma^+$ and $\sigma^-$ circularly polarized light, showing clear valley Zeeman splitting, and the appearance of higher-energy absorption lines corresponding to optical transitions to higher Landau levels ($V_g$=2~V). c,d) The \textit{average} energy of the $\sigma^+$ and $\sigma^-$ absorption lines reveals the diamagnetic shift of the hexciton absorption resonance, which does not follow a purely quadratic dependence (even at low $B$). As a point of reference, the dashed red curves depict a $\sigma B^2$ shift, using a diamagnetic coefficient $\sigma=3~\mu$eV/T$^2$, which is $\approx$10$\times$ larger than the small diamagnetic shifts of the neutral exciton in WSe$_2$ and WS$_2$.}
\label{fig2}
\end{figure}

The ordering of the spin and valley polarized CBs in monolayer WS$_2$ is similar to that of WSe$_2$ (i.e., $\Delta_c$ is also positive), and optical transitions at the A-exciton couple to the upper CBs. Consequently, optical signatures of composite hexcitons can therefore be anticipated at high $n_e$ in WS$_2$ monolayers.  The gate-dependent absorption map of Fig. 1c confirms this prediction: Strong absorption from $X^0$ is plainly visible at $V_g \approx 0$, and $X_{s,t}^-$ charged excitons appear at low $n_e$. These conventional $X_{s,t}^-$ resonances in WS$_2$ have been clearly resolved in several recent studies \cite{Kapuscinski:2020, Zipfel:2020, Zinkiewicz:2021, Robert:2021NC}. Most importantly, our dual-gated structure allows a smooth tuning to a regime of high $n_e$, where Fig. 1c shows that $X_{s,t}^-$ disappear and a new strong absorption resonance emerges at even lower energy ($\approx$15~meV below $X_s^-$). The gate-dependent absorption of monolayer WS$_2$ is therefore qualitatively identical to that of WSe$_2$, albeit with broader linewidths that are likely due to the reduced material quality of sulfur-based TMDs. Thus, we associate the emergence of the low-energy absorption resonance with the stable formation of 6-particle hexciton states.  Note that we were unable to dope our WS$_2$ monolayers with mobile holes; even at large negative $V_g$, only the neutral $X_0$ exciton was visible, likely due to strong mid-gap pinning of the Fermi level by the larger number of defects in sulfur-based TMDs. 

Additional evidence supporting a picture of hexcitons in monolayer WS$_2$ is the evolution of its optical resonance in applied magnetic fields $B$. Figure 2 shows circularly polarized magneto-absorption from both WSe$_2$ and WS$_2$, under conditions of large $n_e$ where the hexciton absorption dominates.  As shown recently \cite{Li:2022}, the hexciton resonance in WSe$_2$ monolayers splits and shifts with increasing $B$, and additional absorption resonances appear at higher energy that disperse linearly with $B$ and are related to the development of Landau levels (LLs) in the conduction and valence bands.  Figure 2 shows that a qualitatively similar $B$-dependent evolution of the hexciton peak also exists in WS$_2$, where additional LL-like absorption features emerge for $B>30$~T.  From the separation of these peaks we can estimate a combined electron and hole cyclotron resonance energy of $\approx$0.45 meV/T, which is slightly larger than obtained from WSe$_2$ \cite{Li:2022} but in line with expectation given the slightly lighter carrier masses in WS$_2$ \cite{Goryca:2019}. 

\begin{figure} 
\center
\includegraphics[width=.45\textwidth]{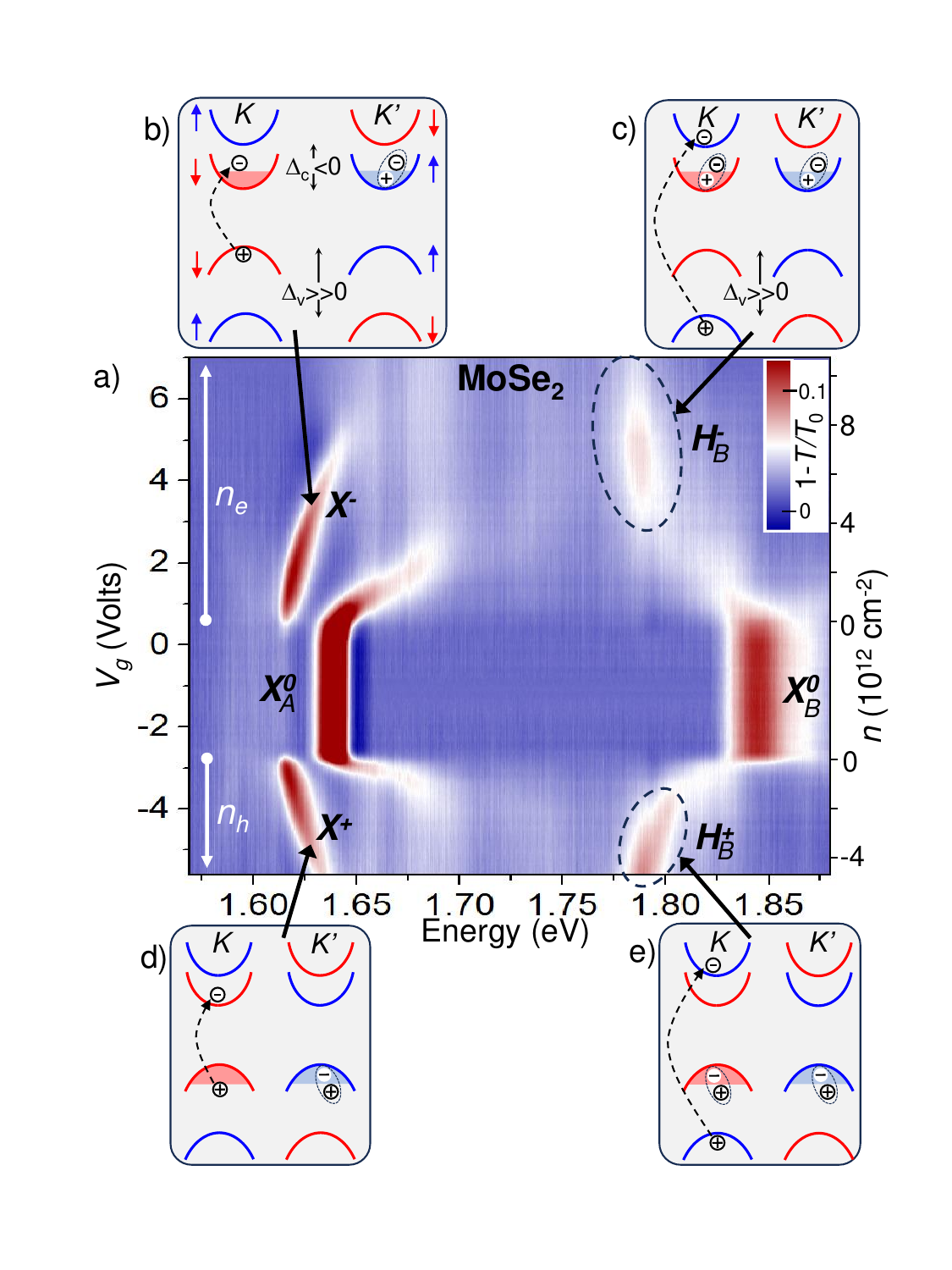}
\caption{a) Gate-dependent absorption of a charge-tunable MoSe$_2$ monolayer at 4~K.  Only conventional $X^-$ and $X^+$ are observed at the A-exciton ($\sim$1.63 eV), even at large electron or hole densities, because $\Delta_c<0$ and therefore only a single distinguishable Fermi sea is available -- see diagrams in panels b) and d), respectively. But at the B-exciton ($\sim$1.83 eV), photoexcited \textit{e-h} pairs couple to the upper CBs, and interaction with both Fermi seas in the lower CBs is possible, leading to hexciton formation at higher $n_e$ -- see diagram in panel c).  Moreover, hexciton formation at high hole densities is always possible at the B-exciton (for \textit{all} TMD monolayers, because $\Delta_v \gg 0$) -- see diagram in panel e). Blue/red colors in the diagrams denote spin-up/down bands.  For clarity, only optical transitions in the $K$ valley are depicted.}
\label{fig3}
\end{figure}

Moreover, from these spectra we can extract the diamagnetic shifts of the hexciton resonance, given by the \textit{average} of the $\sigma^+$ and $\sigma^-$ absorption energies. This analysis, shown in Figs. 2c and 2d, reveals that the diamagnetic shifts of hexcitons -- even at low $B$ -- deviate from the purely quadratic behavior that is known to exist for neutral excitons in WSe$_2$ and WS$_2$ \cite{Stier:2018, Goryca:2019} (especially for WS$_2$, where atypical diamagnetic shifts have also been observed for charged excitons \cite{Plechinger:2016}). This behavior may arise from non-zero angular momentum contributions from the particles in the hexciton complex, and will be further investigated in future studies. For reference, however, the dotted red curves depict purely quadratic shifts ($\sigma B^2$) using a large diamagnetic coefficient $\sigma =3~\mu$eV/T$^2$, which is $10\times$ larger than the known diamagnetic shifts of the very small and tightly-bound neutral excitons \cite{Stier:2018, Goryca:2019}. To the extent that these curves approximately capture the overall shifts of the hexciton resonances, these data are  qualitatively consistent with recent calculations indicating that composite hexcitons are several times larger in size than neutral excitons \cite{vanTuanPRL:2022, vanTuan:2022}

Taken together, these data provide evidence for the emergence of composite hexciton states in electron-doped monolayer WS$_2$, which appear at the A-exciton owing to the positive sign of $\Delta_c$ and consequent ordering of the spin-orbit-split CBs in the $K$ and $K'$ valleys. 

In marked contrast, neither electron-doped nor hole-doped MoSe$_2$ monolayers show any indication of hexciton formation at the A-exciton, as shown in Fig. 3a. Rather -- and as also observed in earlier studies \cite{Wang:2017NL, Smolenski:2019, Liu:2021}-- charge-tunable MoSe$_2$ monolayers exhibit only a single $X^-$ and $X^+$ absorption that appears $\approx$25~meV below the neutral exciton $X_A^0$.  This observation is consistent with the negative sign of $\Delta_c$ in MoSe$_2$ \cite{Song:2013, Kormanyos:2015}, which dictates that optical transitions at the A-exciton photoexcite electrons to the lower (not upper) CBs.  As such, when mobile electrons populate the lower CBs, only a single type of distinguishable electron exists in the Fermi sea, and many-body hexcitons cannot form (see diagrams in Fig. 3b,d).  

The ordering of the CBs in MoSe$_2$ \textit{does}, however, permit hexciton formation at the higher energy B-exciton, where optical transitions couple to the upper CBs (see Fig. 3c).  As Fig. 3a shows, when $n_e$ is small, the neutral B-exciton ($X_B^0$) at 1.84~eV disappears and only a faint and diffuse absorption remains at energies where conventional charged excitons are expected (that is, at about 25~meV below $X_B^0$, or $\approx$1.815~eV).  However, at larger $n_e$ ($\approx 5 \times 10^{12}$~cm$^{-2}$), a stronger absorption with increased oscillator strength clearly emerges at $\approx$1.795~eV.  Its separation from $X_B^0$ is $\approx$45~meV, which is larger than the 25~meV expected for conventional charged excitons, but is commensurate with the larger energy separation between hexcitons and neutral excitons that we observed in WSe$_2$ and WS$_2$ (\textit{cf.} Fig. 1). Notably, Fig. 3 shows that this new absorption does not emerge \textit{until} a large $n_e$ comparable to where the conventional $X^-$ trion loses oscillator strength, suggesting that it is not a conventional trion.  Therefore, although past studies of charge-tunable MoSe$_2$ monolayers have associated similar spectral signatures with conventional trions/tetrons/exciton-polarons of the B-exciton \cite{Wang:2017NL, Liu:2021}, we argue that the low energy of the emerging absorption resonance and (especially) its dependence on $n_e$ are, in fact, more consistent with the formation and emergence of many-body hexcitons in electron-doped MoSe$_2$ monolayers.

Furthermore, analogous signatures of hexcitons appear when the MoSe$_2$ monolayer is doped with high concentrations of mobile holes, $n_h$ (see Fig. 3a). As depicted in the diagram of Fig. 3e, \textit{e-h} pairs excited at the B-exciton will \textit{always} have two distinguishable Fermi seas of mobile holes with which to interact, and composite hexcitons can be anticipated. Figure 3a shows that as $n_h$ starts to increase, both $X_A^0$ and $X_B^0$ disappear, and a conventional $X^+$ resonance appears 25~meV below $X_A^0$ when $V_g =-3$~V.  However, the concomitant response at the B-exciton is much weaker, until $V_g$ increases up to $\approx -4$~V, at which point a strong absorption emerges at a much lower energy of $\approx$40 meV below $X_B^0$. This new resonance actually \textit{gains} oscillator strength with increasing $n_h$, while in this same doping range the conventional $X^+$ fades away. Based on this $n_h$ dependence, we rule out the possibility that the new resonance could be, e.g., a conventional trion associated with $X_B^0$ or an excited Rydberg state of $X_A^0$. Moreover, this new resonance redshifts with increasing $n_h$, similar to the redshift observed for hexcitons in electron-doped WSe$_2$ and WS$_2$ (cf. Fig. 1). These data therefore support a picture of robust hexciton formation at the B-exciton in hole-doped MoSe$_2$. The resonance features are rather broad, however, likely due  to the shorter lifetime of B-excitons.  

Indeed, owing to the large 100s-of-meV spin-orbit splitting of the valence bands ($\Delta_v$) that exists in all monolayer TMD semiconductors, robust hexciton formation at the B-exciton should emerge when \textit{any} TMD monolayer semiconductor is doped with a high density of mobile holes (see diagram in Fig. 3e). Furthermore, because $\Delta_v$ is large, the Fermi sea of holes never occupies the valence bands from which the photoexcited \textit{e-h} pair originates, and the photohole always remains distinguishable from every hole in the Fermi sea. Hexcitons under such conditions should therefore remain robust and should redshift due to the primary effects of bandgap renormalization \cite{Scharf:2019} up to very large values of $n_h$.  While in this work we have studied B-excitons only in MoSe$_2$, we note that the seminal work of Wang et al. \cite{Wang:2017NL} clearly revealed a strong absorption below $X_B^0$ in heavily hole-doped WSe$_2$ monolayers, that redshifted with increasing $n_h$ and did not broaden, consistent with hexciton formation. 

An important insight gained from these various results is that the decay and broadening of excitonic complexes is likely governed more by the quantum-mechanical distinguishability of the photoexcited \textit{e-h} pair (in relation to the carriers in the Fermi sea), than by screening from the Fermi sea. For example, in the case of electron-doped WSe$_2$ and WS$_2$ shown in Fig. 1, the redshifts of the hexciton resonances in WSe$_2$ and WS$_2$ cease, and their decay/broadening begins, only when electrons begin to populate the upper CBs. Beyond this point, the photoexcited electron is no longer distinguishable from every electron in the Fermi sea, and the Pauli exclusion principle dictates that when it is introduced into the (now occupied) upper CB, it must scatter away those mobile electrons having similar spin and valley quantum numbers \cite{vanTuan:2022}. This exchange scattering process leads to the decay and broadening of the hexciton resonance. Hexciton resonances therefore retain their amplitude and narrow linewidth when the charge density increases, as long as the photoexcited \textit{e-h} pair remains distinguishable from all carriers in the Fermi sea. In hole-doped TMD monolayers, the large $\Delta_v$ ensures distinguishability of \textit{e-h} pairs excited at the B-exciton, and therefore a robust stability of hexcitons even to very large $n_h$. 

Moreover, screening by mobile carriers does not seem to be the primary driving force for broadening of the hexciton resonance, as inferred by the observation that the conventional $X^\pm$ resonances begin to lose oscillator strength at much smaller carrier density. For example, Fig. 3 shows that the conventional trions of MoSe$_2$ around 1.63~eV decay when $|V_g| \gtrsim 5$~V. Concomitantly, however, the type-B hexciton in hole-doped conditions around 1.8~eV neither broadens nor decays at the same (and even higher) $n_h$. A similar behavior was also measured for a wider $n_h$ range by Liu et al. \cite{Liu:2021}, where the type-B resonance around 1.8~eV maintains a large oscillator strength and continues to redshift long after the $X^+$ resonance at the A-exciton decays. This suggests that screening is not the primary cause of decay and broadening, because the Coulomb potential cannot selectively weaken the attraction between particles of one complex species but not of another. 

In summary, composite hexcitons are expected in \textit{all} members of the monolayer TMD family. When doped with a high density of electrons, hexcitons can emerge at the A-exciton resonance (as for the case of WSe$_2$ and WS$_2$) or the B-exciton resonance (as for the case of MoSe$_2$) depending on the sign of $\Delta_c$ and the consequent ordering of the CBs. For hole-doped TMD monolayers, hexcitons should \textit{always} emerge at the B-exciton, as investigated here for MoSe$_2$ and as suggested by earlier spectroscopic data from both WSe$_2$ and MoSe$_2$ \cite{Wang:2017NL, Liu:2021}. To the best of our knowledge, this has not yet been studied in hole-doped WS$_2$ or MoS$_2$. As a final point of discussion, we note that electron-doped MoS$_2$ monolayers represent an interesting case: While early theory suggested that $\Delta_c$ was small and negative (implying a CB ordering similar to MoSe$_2$ \cite{Kormanyos:2015}), more recent experimental work indicates an opposite CB ordering \cite{Robert:2020, Park:2022}, particularly when exciton effects are taken into account, making it more akin to that of WS$_2$ and WSe$_2$.  In this case, a hexciton resonance can be expected to emerge in electron-doped MoS$_2$ at the A-exciton, at an energy below that of the conventional $X_{s,t}^-$ charged excitons.  Indeed, various recent studies have revealed additional optical resonances emerging in electron-doped MoS$_2$ monolayers \cite{Klein:2021, Klein2:2021, Roch:2019}, although in some cases it was associated with exotic ferromagnetic order.  Looking forward, we anticipate that composite 6-particle hexcitons (and, in high magnetic fields, 8-particle oxcitons) can provide a rich platform to study novel many-body effects and inter-valley correlations within a strongly interacting exciton-Fermi sea. 

We thank X. Marie for helpful discussions, and we acknowledge support from the Los Alamos LDRD program and the US Department of Energy (DOE) ``Science of 100~T'' program. The National High Magnetic Field Lab is supported by National Science Foundation DMR-1644779, the State of Florida, and the US DOE. Work at the University of Rochester was supported by the DOE Basic Energy Sciences, Division of Materials Sciences and Engineering under Award No. DE-SC0014349.

%\bibliographystyle{apsrev4-1}
%\bibliography{ASI_lattices3}

\end{document}